\documentclass{IEEE_lsens}
\usepackage{textcomp}
\usepackage[noadjust]{cite}
\ifCLASSINFOpdf
\else
\fi
\usepackage[T1]{fontenc} 
\usepackage{amsmath}
\usepackage{graphicx}
\usepackage{subfigure}
\usepackage{caption}
\usepackage{subcaption}
\usepackage{xcolor}
\usepackage[cmintegrals]{newtxmath}
\usepackage{bm}
\usepackage{array}
\usepackage{url}

\ifCLASSINFOpdf
\else
\fi
\providecommand{\hypersetup}[1]{\relax}

\hypersetup{pdftitle={Bare Demo of IEEE\_lsens.cls for IEEE Sensors Letters},
pdfsubject={Typesetting},
pdfauthor={Michael D. Shell},
pdfkeywords={Class, IEEE, IEEE\_lsens, IEEE Sensors Letters, LaTeX, Typesetting, TeX}}
\begin{document}
\markboth{Vol.~1, No.~3, July~2017}{0000000}
\IEEELSENSarticlesubject{Sensor Applications}

\title{In-Orbit Cosmo-SkyMed antenna pattern estimation by a narrowband sweeper receiver}

\author{\IEEEauthorblockN{Mohammad~Roueinfar\IEEEauthorrefmark{1}, 
and~Masoud~Ardini\IEEEauthorrefmark{2}}
\IEEEauthorblockA{\IEEEauthorrefmark{1}School of Electrical Engineering, Iran University of Science and Technology, Tehran, Iran\\
\IEEEauthorrefmark{2}School of Electrical Engineering, Shahed University, Tehran, Iran\\
\\
}
\thanks{Corresponding author: M. Roueinfar (e-mail: m\_roueinfar@alumni.iust.ac.ir).\protect\\
(For the real e-mail address, see http://www.michaelshell.org/contact.html).}
\thanks{Associate Editor: Alan Smithee.}%
\thanks{Digital Object Identifier 10.1109/LSENS.2017.0000000}}
\IEEELSENSmanuscriptreceived{Manuscript received June 7, 2017;
revised June 21, 2017; accepted July 6, 2017.
Date of publication July 12, 2017; date of current version July 12, 2017.}

\IEEEtitleabstractindextext{
\begin{abstract}
This paper introduces a novel method for antenna pattern estimation in satellites equipped with Synthetic Aperture Radar (SAR), utilizing a Narrowband Sweeper Receiver (NSR). By accurately measuring power across individual frequencies within SAR’s inherently broadband spectrum, the NSR significantly enhances antenna pattern extraction accuracy. Analytical models and practical experiments conducted using the Cosmo-SkyMed satellite validate the receiver’s performance, demonstrating superior signal-to-noise ratio (SNR) compared to conventional receivers. This research represents a key advancement in SAR technology, offering a robust framework for future satellite calibration and verification methodologies.
\end{abstract}

}

\IEEEpubid{1949-307X \copyright\ 2017 IEEE. Personal use is permitted, but republication/redistribution requires IEEE permission.\\
See \url{http://www.ieee.org/publications\_standards/publications/rights/index.html} for more information.}

\maketitle

\section{Introduction}
The use of Synthetic Aperture Radar (SAR) in satellite systems has revolutionized remote sensing, enabling high-resolution imagery across various application, including environmental monitoring, disaster management and urban planning. The accurate estimation of the SAR antenna pattern is essential for optimizing image quality and ensuring the reliability of SAR data. Traditional methods for estimating the antenna pattern often rely on ground-based receivers, which present limitations in terms of spatial coverage and measurement accuracy \cite{1}. However, recent advancements in receiver technology have facilitated the development of more sophisticated approaches to antenna pattern estimation.
\\
Calibration and verification methods for SAR antennas have been described in \cite{2}. Measuring and estimating the SAR antenna pattern after satellite launch is one of the key techniques. It is possible to vary the phase and amplitude of the excitation coefficients to adjust the antenna pattern \cite{3}. Ground-based receivers can be used to estimate the SAR antenna pattern, and the estimated antenna pattern is used to correct the excitation coefficients. Moreover, these reference patterns are used to eliminate the pattern shape from the SAR images \cite{3}. In \cite{4}, \cite{5}, some receivers for estimating the SAR antenna pattern are described, all of which use a Simple Envelope Detector (SED) to extract and process the video signal. The azimuth pattern can be recorded with one receiver, while the elevation pattern requires multiple receivers that need to be placed in the specified areas \cite{6}. The challenges of measuring the azimuth antenna pattern in a bistatic interferometry SAR (InSAR) configuration, especially due to overlapping signals from two satellites, are addressed in \cite{7}.\\
A novel approach for extracting the active phased array antenna pattern of the in-orbit KOMPSAT-5 satellite utilizing T/R module calibration data is presented in \cite{8}. This advancement underscores the importance of calibration data in enhancing the accuracy of antenna pattern measurements. 
Reference \cite{9} discusses a calibrator for low-frequency SAR point targets based on radar cross-section and temporal stability. By analyzing an interferometric stack of SAR acquisitions, stable point targets can be identified and subsequently used as passive calibrators to validate radiometry, antenna elevation patterns, and pointing accuracy.
The principles, design, construction, and measurement results of a new Polarimetric Active Radar Calibrator (PARC) designed for the GF-3 satellite are detailed in \cite{10}. This includes a new design methodology for PARC, which employs two dual-polarized antennas characterized by exceptionally high polarization purity. 
A new method for measuring the azimuth pointing of spaceborne SAR antenna beams is proposed in \cite{11}. This method utilizes a ground receiver capable of receiving and recording complex sampling data from pulse signals transmitted by the GaoFen-3 satellite.
A space-based method for in-orbit measurement of the MEO-SAR elevation antenna pattern, utilizing a nano-calibration satellite equipped with a receiver, is proposed in \cite{12}. This innovative approach enables the nano-calibration satellite to fly across the entire MEO-SAR swath, allowing for the extraction of elevation antenna pattern coverage from the recorded data.
An effective method for estimating the elevation antenna pattern (EAP) of the SuperView Neo-2 SAR satellite is proposed in \cite{13}. By leveraging in-orbit measurements and natural features such as the Amazon rainforest, this method achieves precise calibration.\\
In this paper, we present the following key contributions. First, we introduce  the Narrowband Sweeper Receiver (NSR) for SAR satellite antenna pattern estimation. The NSR is capable of measuring power at each frequency within the SAR signal spectrum. This capability
is particularly significant given the broadband nature of SAR signals, which often exhibit variations in antenna patterns across different frequencies. Additionally, the NSR not only enhances the precision\\
\\
\\
of antenna pattern extraction but also improves the signal-to-noise ratio (SNR), enabling more accurate data interpretation. The rest of the paper is organized as follows. In Section II, the proposed receiver is discussed, detailing its design and functionality in antenna pattern estimation. Section III presents the practical results obtained from the NSR, highlighting its performance in extracting antenna patterns at various frequencies. Finally, the conclusion is summarized in Section IV.
\section{SAR satellite antenna}
The antenna mounted on the satellite utilizes the phased array structures. In \cite{2}, \cite{3}, the antenna pattern equation is provided, where $\varepsilon$ and $\alpha$ represent the elevation and azimuth angles, respectively. The number of antennas in this array is denoted as $N$ out of $M$. Additionally, $\Delta x$ and $\Delta y$ define the width and height of the antenna, respectively. In this equation, $k_{0}$ is the wave number, calculated using $k_{0} = 2\pi / \lambda$ .The term $\vec{\textbf{C}}_{SA,mn}$ describes the radiation characteristics of each element, and $ \textbf{E}_{SA,mn} $  constitutes the error matrix.
\begin{align}
\label{e1}
\vec{\textbf{F}}_{Beam}(\varepsilon,\alpha)=\sum_{m=0}^{M-1}  \sum_{n=0}^{N-1} \Big( \vec{\textbf{C}}_{SA,mn}(\varepsilon,\alpha)  \alpha_{mn}  \textbf{E}_{SA,mn}  \nonumber \\
\cdot \exp (jk_{0} \sin(\varepsilon) \cos(\alpha) \big(- \frac{N-1}{2} \big) \Delta y   \nonumber \\
\cdot \exp (jk_{0} \cos(\varepsilon) \sin(\alpha) \big(- \frac{M-1}{2} \big) \Delta x \Big)
\end{align}
The range resolution is achieved by (\ref{e2}) \cite{14}. In this equation,, $c$ is the speed of light, $BW$ is bandwidth, $\delta$ is the range resolution, and $\psi$ is the incidence angle.
\begin{equation}
\label{e2}
\delta \cong \frac{c}{2 \cdot BW  \cdot \sin(\psi)}
\end{equation}
The frequency response of the antenna will lack sufficient flatness if the antenna's pattern changes significantly with frequency variations. Consequently, the brightness of image pixels depends on the target radar cross-section (RCS) and antenna flatness. The undesirable effect on antenna flatness can be mitigated by measuring the antenna shape at each frequency. The receiver used in this research can extract the antenna pattern at each frequency individually. Furthermore, the implemented receiver can improve the signal-to-noise ratio (SNR) for antenna pattern extraction.
\section{The Proposed Narrowband Sweeper Receiver (NSR)}
\subsection{Narrowband Sweeper Receiver Performance}
\begin{figure}[h]
\centering
\includegraphics[width=3in]{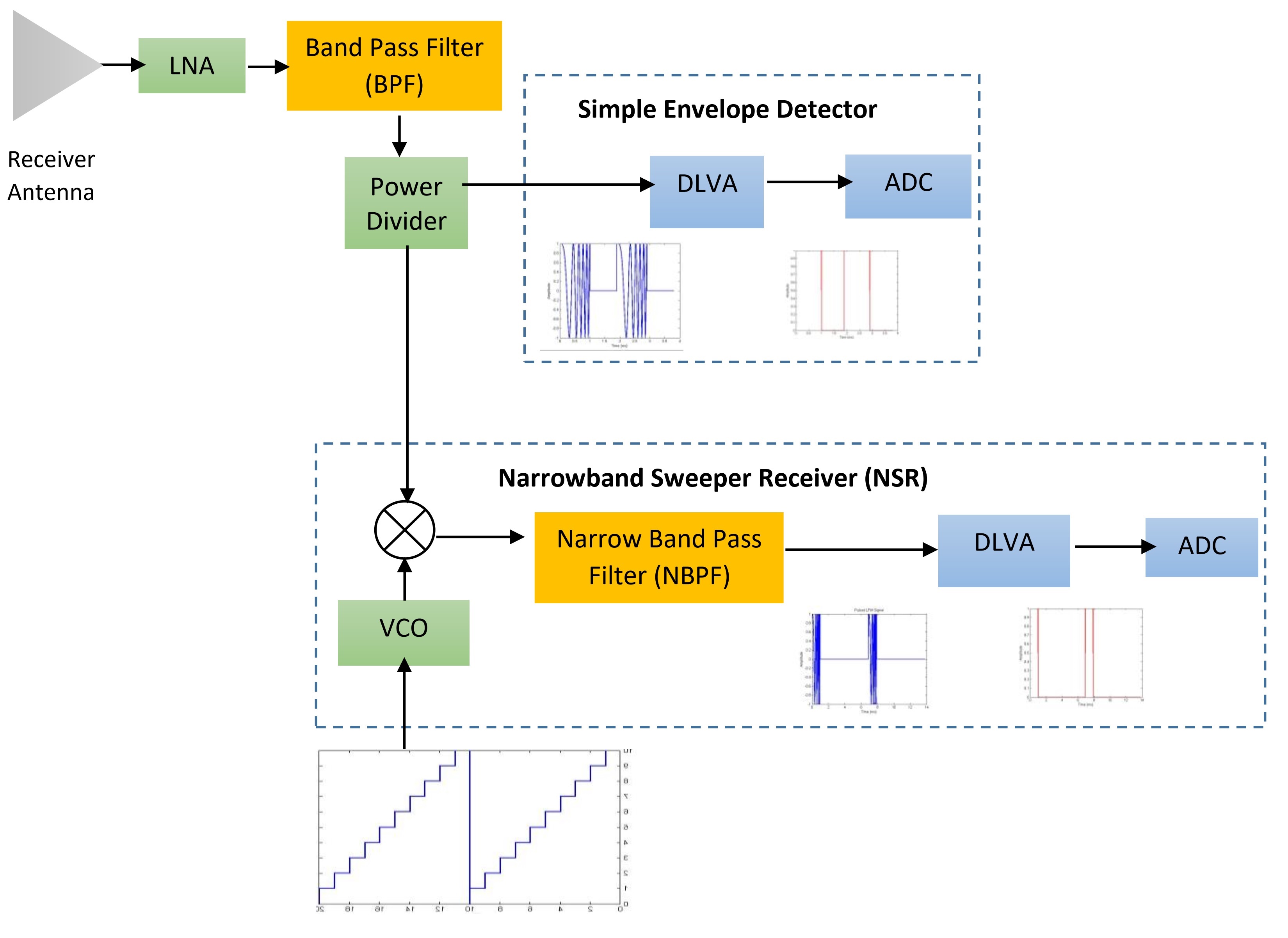}
\caption{The proposed Receiver Based on Narrowband Sweeper Receiver (NSR).}
\label{Fig1}
\end{figure}

\begin{figure}
\centering
\includegraphics[width=1.8in]{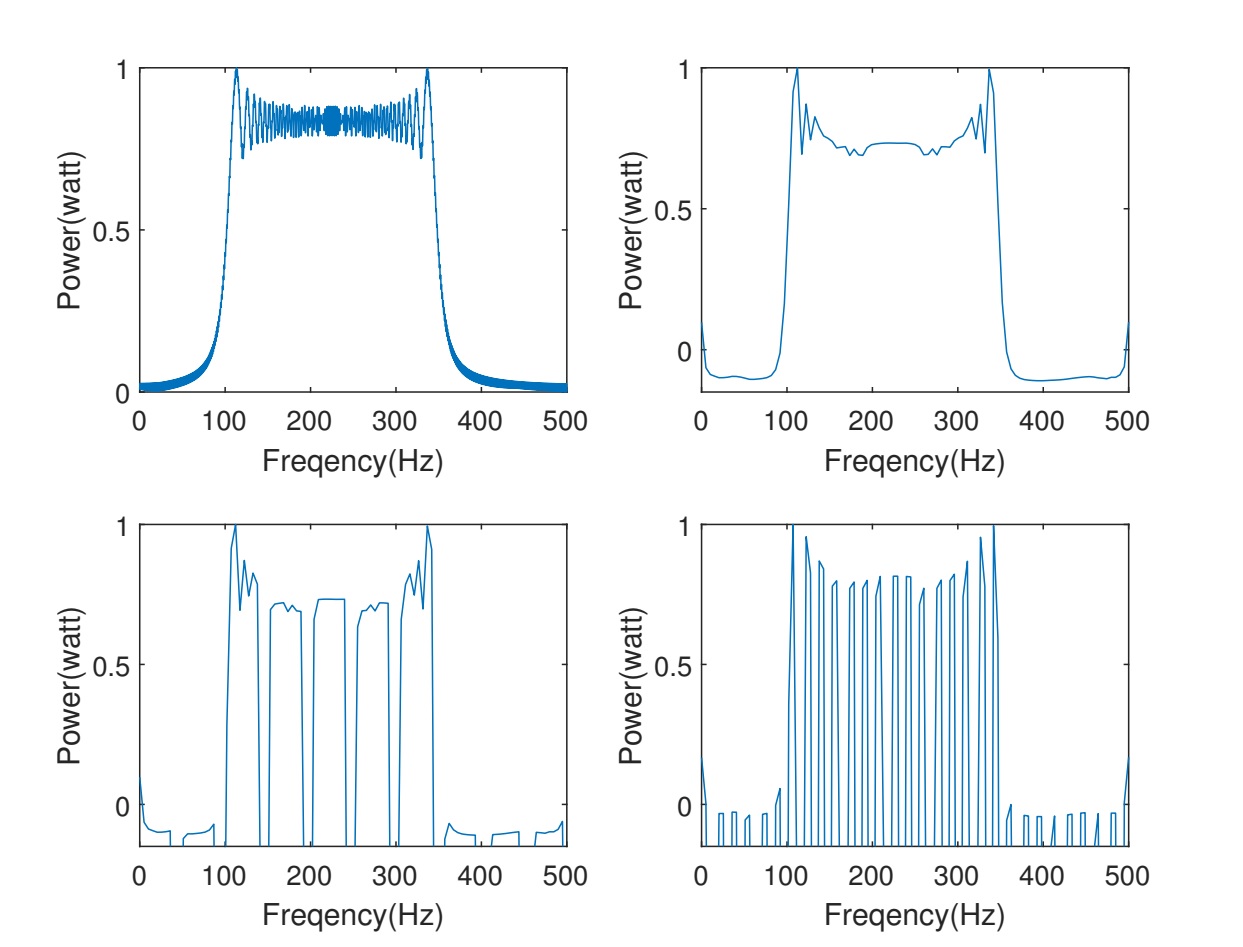}
\caption{a) The original spectrum, b) The reconstructed spectrum when $T_{stop}=2PRI$, c) The reconstructed spectrum when $T_{stop}=0.9 PRI$.}
\label{Fig2}
\end{figure}

Fig.\ref{Fig1} illustrates the structure implemented in this research. In addition to implementing the Narrowband Sweeper Receiver (NSR), the receiver described in \cite{4} and \cite{5} was also implemented for comparative analysis. In this configuration, the SAR satellite signal as a pulsed linear frequency modulated (LFM) signal, is received by the antenna, amplified by a low-noise amplifier (LNA), and passed through a bandpass filter (BPF). To compare the performance of the simple envelope detector (SED) and NSR receivers, the signal is divided using a power divider, enabling parallel input to both receivers. In the SED receiver, the received signal is demodulated using a detector logarithmic video amplifier (DLVA) and converted into digital format through an analog-to-digital converter (ADC) for further processing and parameter extraction. Simultaneously, a copy of the signal is routed to the NSR receiver, where it undergoes mixing with a voltage-controlled oscillator (VCO) signal driven by a ramp input. The VCO output is filtered through a narrowband bandpass filter (NBPF), demodulated using a DLVA, and digitized via an ADC for subsequent analysis.
The proposed structure in Fig.\ref{Fig1} is equivalent to an NBPF moving through the spectrum. The NSR consisted of a swept narrowband filter in the frequency spectrum that can measure the power of signal at the desired frequency at each stop. Additionally, this structure can measure the power at each frequency, as well as the shape of the spectrum, central frequency, and bandwidth. The estimation accuracy of the central frequency and bandwidth depends on the bandwidth of the NBPF. The period of ramp signal $T$ indicates the sweep speed and  $T_{stop}$ indicates the stair step. The NBPF will have a 3dB overlap in the subsequent steps if the receiver's parameters are configured as per (\ref{e3}).
\begin{equation}
\label{e3}
T_{stop}=\frac{NBPF}{BPF} \times T
\end{equation}
The spectrum shape would be incorrect if the pulse repetition interval (PRI) of the received signal is smaller than $T_{stop}$. In this case, the spectrum shape would appear as several sequential impulses, as no pulse is received at certain stop times. The condition for accurate estimation is shown in (\ref{e4}).
\begin{equation}
\label{e4}
T_{stop} < PRI
\end{equation}
and we have
\begin{equation}
\label{e5}
\frac{NBPF}{BPF} \times T < PRI	
\end{equation}
The simulation results are presented in Fig. \ref{Fig2}. LFM signals with a 20\% duty cycle are used as the input for the NSR. Fig. \ref{Fig2}-a shows the original spectrum, Fig. \ref{Fig2}-b demonstrates the estimated spectrum when $T_{stop}=2 PRI$, and Fig. \ref{Fig2}-c shows the estimated spectrum when $T_{stop}=0.9 PRI$.
\subsection{Effect of Frequency Variation on Antenna Pattern}	
\begin{figure}[h]
\centering
\includegraphics[width=1.6in]{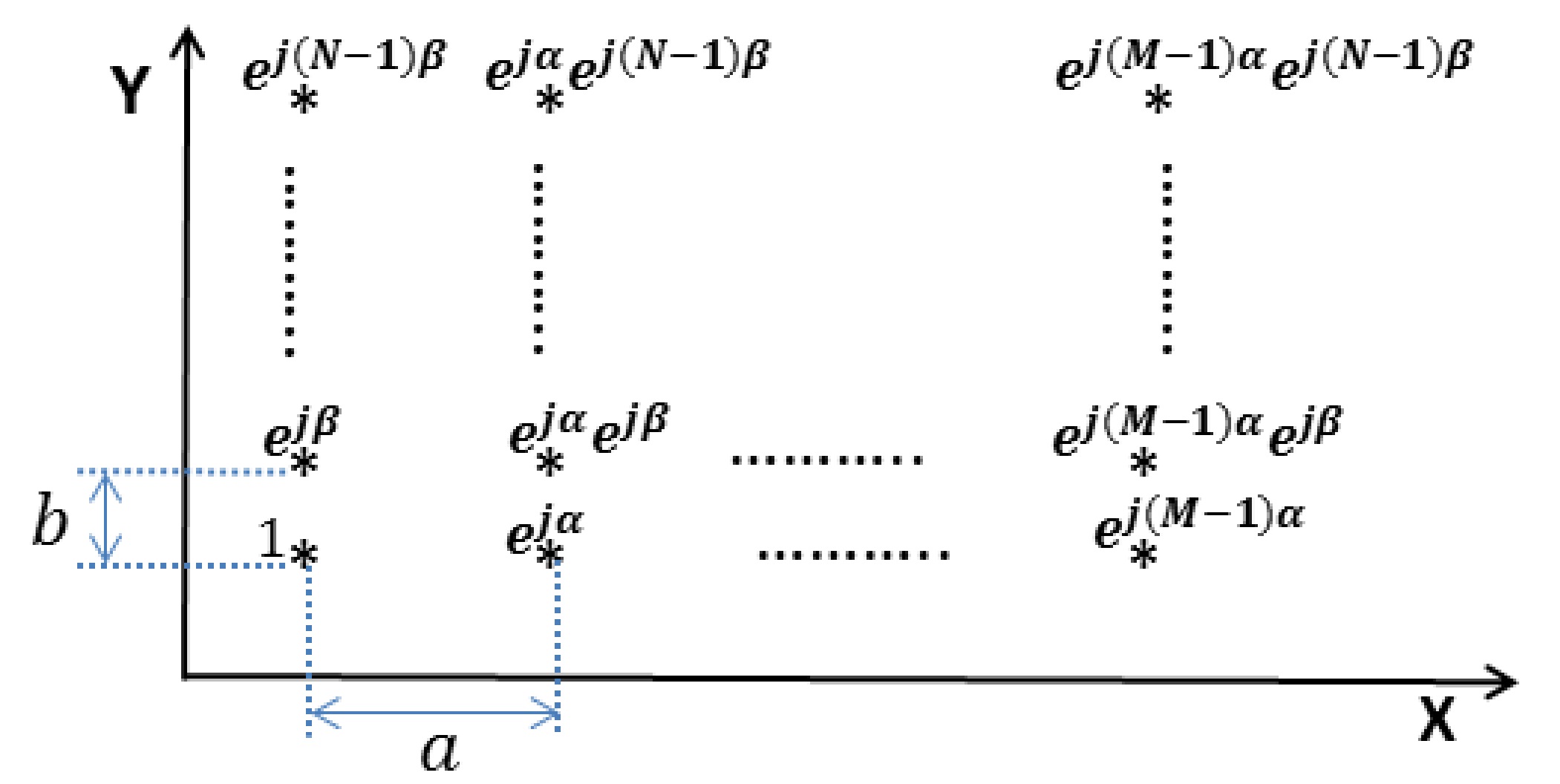}
\caption{The electric current excited of the elements.}
\label{Fig3}
\end{figure}

\begin{figure}[h]
\centering
\includegraphics[width=2in]{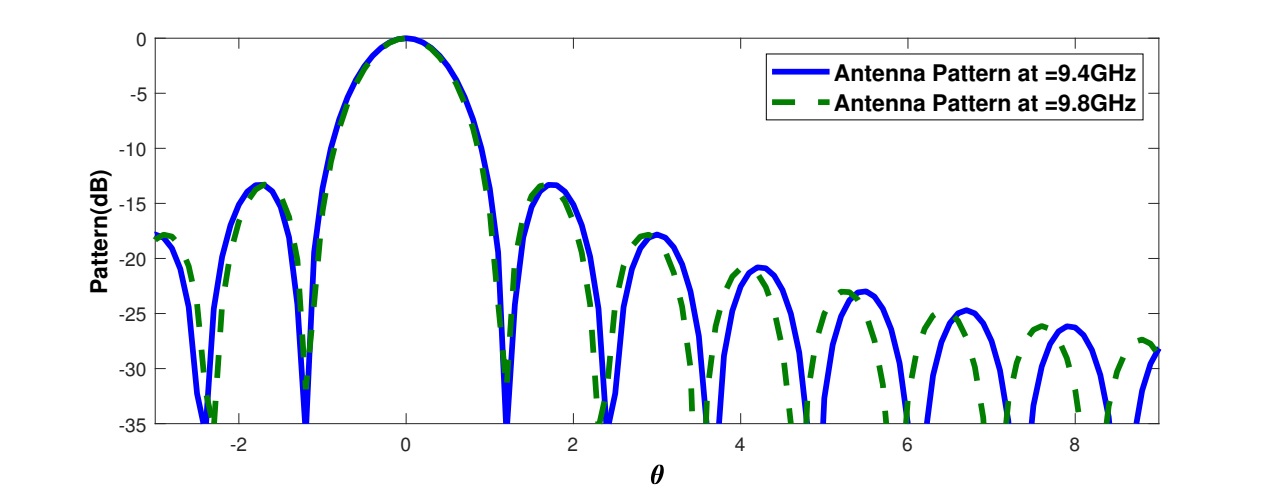}
\caption{The antenna pattern at two different frequencies.}
\label{Fig4}
\end{figure}

The NSR can measure signal power across all frequencies. In this section, we aim to identify both anticipated and unanticipated behaviors in the antenna pattern. The Cosmo-SkyMed SAR instrument features an active phased array antenna \cite{15}. The antenna pattern is derived by multiplying the array factor (AF) with the element factor (EF).
\begin{equation}
\label{e6}
F(\theta,\varphi)= EF(\theta,\varphi) \cdot AF(\theta,\varphi)
\end{equation}
A simple phased array antenna is considered as a model for Cosmo-SkyMed antenna until the effect of frequency variations on the AF is reviewed. The elements that are excited electric current are shown in Fig. \ref{Fig3}. The beam direction can change with variations in $\alpha$ and $\beta$. In this figure, $a$ and $b$ represent the distances between elements.
The $AF$ expression is given in (\ref{e7}) \cite{16}, where $M$ and $N$ are the number of elements along each axis. $I_{mn}$ and $\vec{\textbf{r}}_{mn}$ represent the excitation current and the position vector per element, respectively. Additionally, $k_{0}$ is the wave number, calculated as $k_{0}=\frac{2 \pi}{\lambda}$ \cite{16}.
\begin{equation}
\label{e7}
AF=\sum_{m=0}^{M-1}  \sum_{n=0}^{N-1} I_{mn} \cdot \exp(jk_{0} \hat{\textbf{r}} \cdot \vec{\textbf{r}}_{mn})
\end{equation}
where  $I_{mn}, \vec{\textbf{r}}_{mn}$  and  $\hat{\textbf{r}}  \cdot \vec{\textbf{r}}_{mn}$  are calculated using (\ref{e8}), (\ref{e9}), and (\ref{e10}), respectively.
\begin{equation}
\label{e8}
I_{mn} = \exp(j(m-1) \alpha)  \exp(j(n-1) \beta)
\end{equation}
and
\begin{equation}
\label{e9}
\vec{\textbf{r}}_{mn}= (m-1) a \hat(\textbf{x}) +(n-1) b \hat(\textbf{y}) 
\end{equation}
and
\begin{equation}
\label{e10}
\hat{\textbf{r}} \cdot \vec{\textbf{r}}_{mn}= (m-1) a \sin(\theta) \cos(\varphi) +(n-1) b \sin(\theta) \sin(\varphi)
\end{equation}
(\ref{e11})  is obtained by substituting (\ref{e8})  and (\ref{e10}) into (\ref{e7}) :
\begin{align}
\label{e11}
AF(\theta,\varphi)=  \nonumber \\
&  \sum_{m=0}^{M-1} \exp(j(m-1) (k_{0}a \sin(\theta) \cos(\varphi)+\alpha)) \nonumber \\
&  \sum_{n=0}^{N-1} \exp(j(n-1) (k_{0}b \sin(\theta) \sin(\varphi)+\beta))  \nonumber \\
\end{align}
The normalized antenna pattern is expressed by (\ref{e12}) assuming $\varphi=0$.
\begin{equation}
\label{e12}
|E_{\theta}|= \Bigg| \frac{\sin( \frac{M}{2} k_{0} a \sin(\theta)+\alpha)} {M\sin( \frac{k_{0}a}{2}  \sin(\theta)+\alpha)}  \Bigg|
\end{equation}
The null position in the The antenna pattern at two different frequencies can be calculated if the zeros of (\ref{e13}) are determined.
\begin{equation}
\label{e13}
\sin( \frac{M}{2} k_{0} a \sin(\theta)+\alpha)=0  
\end{equation}
by solving (\ref{e13}) we have $\frac{Ma}{\lambda}\sin(\theta) \times \pi + \acute{\alpha} \times \pi = k \pi$  where $k=0, \pm1 ,\pm2, \dots$ and  (\ref{e14}) is defined if:
\begin{equation}
\label{e14}
\theta= \arcsin (\lambda \times \frac{k-\acute{\alpha}}{Ma}) \qquad k=0, \pm1 ,\pm2, \dots
\end{equation}
According to (\ref{e14}), the distance between the nulls increases proportionally with the wavelength. The array factor pattern at two different frequencies is illustrated in Fig. \ref{Fig4}.
The number of elements and the spacing between elements are set to 100 and 1.5 cm, respectively, in this simulation. As a result, Cosmo-SkyMed's antenna and NSR exhibit correct performance if the behavioral description is observed in practical implementations.
\subsection{Effect of NarrowBand Filter on Antenna Pattern Extraction}
\begin{figure}[h]
\centering
\includegraphics[width=1.3in]{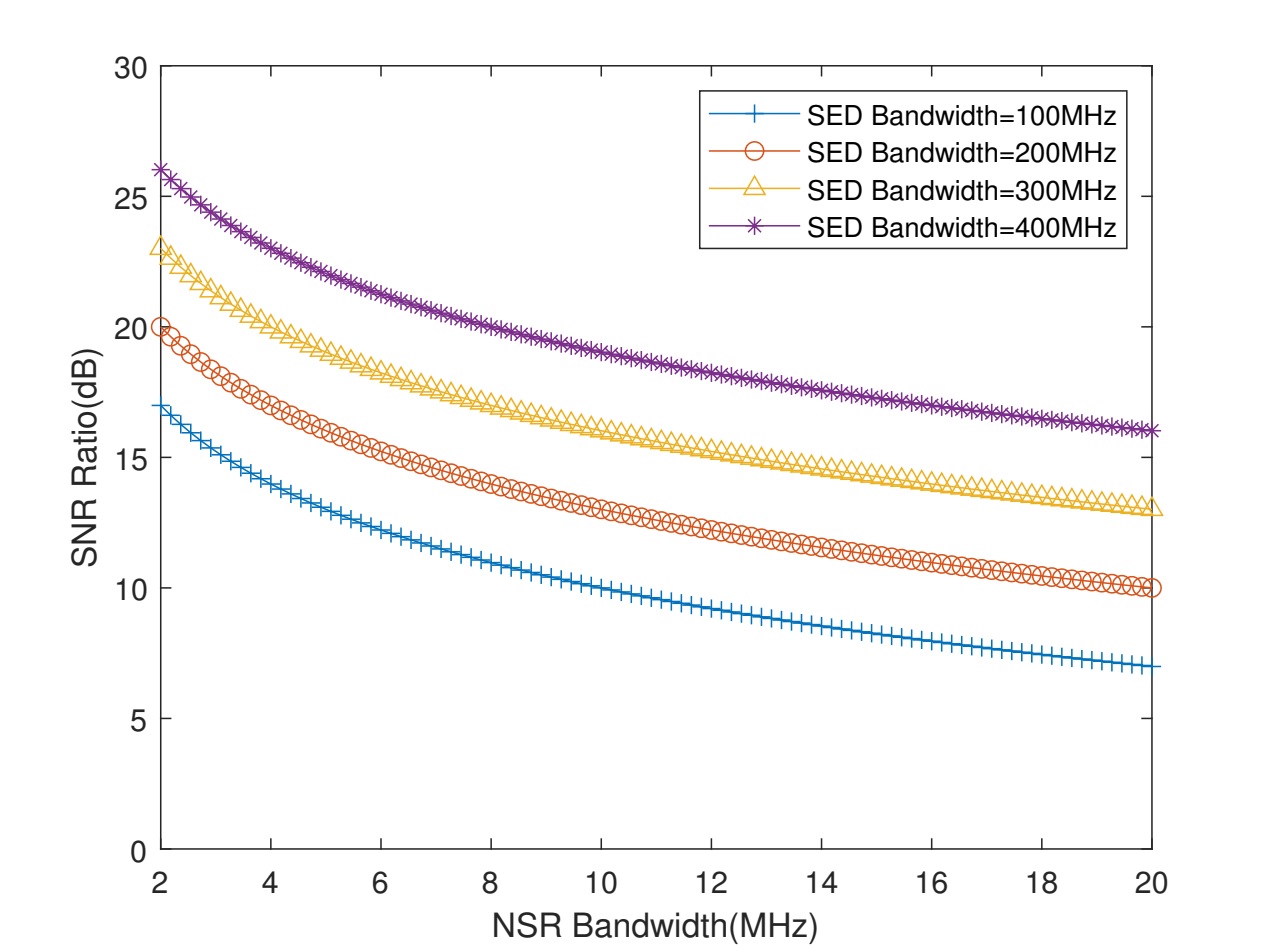}
\caption{SNR ratios illustrated at varying bandwidths.}
\label{Fig5}
\end{figure}

The precision of antenna pattern calibration and image correction depends on the ground receiver's performance. In this section, we review the reasons why NSR is more accurate than SED. (\ref{e15}) shows the SNR of the received signal at the output of the ground receiver's antenna \cite{14}:
\begin{equation}
\label{e15}
SNR= \frac{P_{t}G_{t}G_{r}\lambda^{2}}{(4\pi R)^{2} K T BW}
\end{equation}
where $R$ is the distance from the satellite to the receiver, $BW$ is the bandwidth of the receiver, $P_t$ is the power transmitted of the satellite, $\lambda$ is the wavelength, $K$ is the Boltzmann constant, $T$ is the noise factor, and $G_t$ and $G_r$ are the gains of the satellite's transmitter and ground receiver antennas, respectively.
According to Fig. \ref{Fig1}, the receiver bandwidths are defined by the BPF for SED and the NBPF for NSR. The SNR is represented in (\ref{e16}).
\begin{equation}
\label{e16}
\frac{SNR_{NSR}}{SNR_{SPD}}=\frac{BPF}{NBPF}
\end{equation}
Fig. \ref{Fig5} illustrates the SNR ratio at different receiver bandwidths. NSR uses a filter with a reduced bandwidth than SED. As a result, the SNR of the NSR (compared to SED) achieves an improved equivalent ratio, making NSR more accurate and reliable SED.
\section{Cosmo-SkyMed Satellite pattern estimation using NSR}
\begin{figure}[h]
\centering
\includegraphics[width=2.1in]{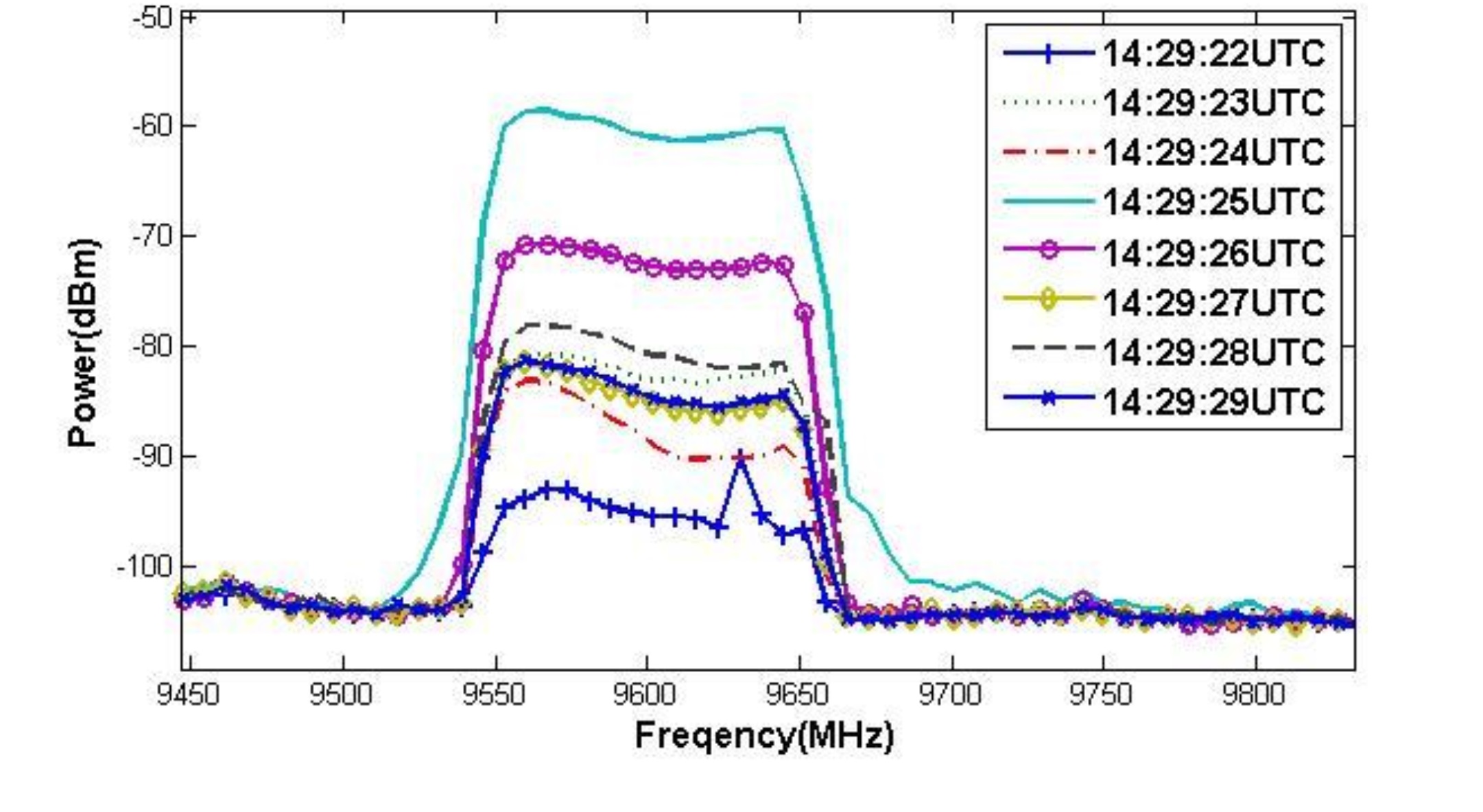}
\caption{Spectrum of the monitored signal at different times.}
\label{Fig6}
\end{figure}

\begin{figure}[h]
\centering
\includegraphics[width=2.4in]{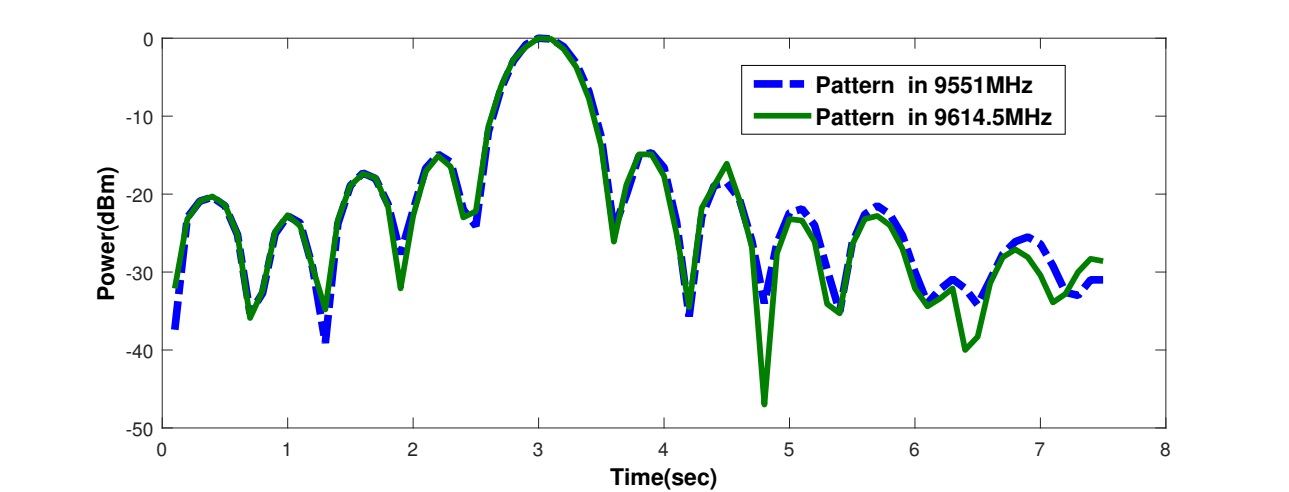}
\caption{Antenna pattern at 9551 MHz and 9614.5 MHz.}
\label{Fig7}
\end{figure}

\begin{figure}[h]
\centering
\includegraphics[width=1.8in]{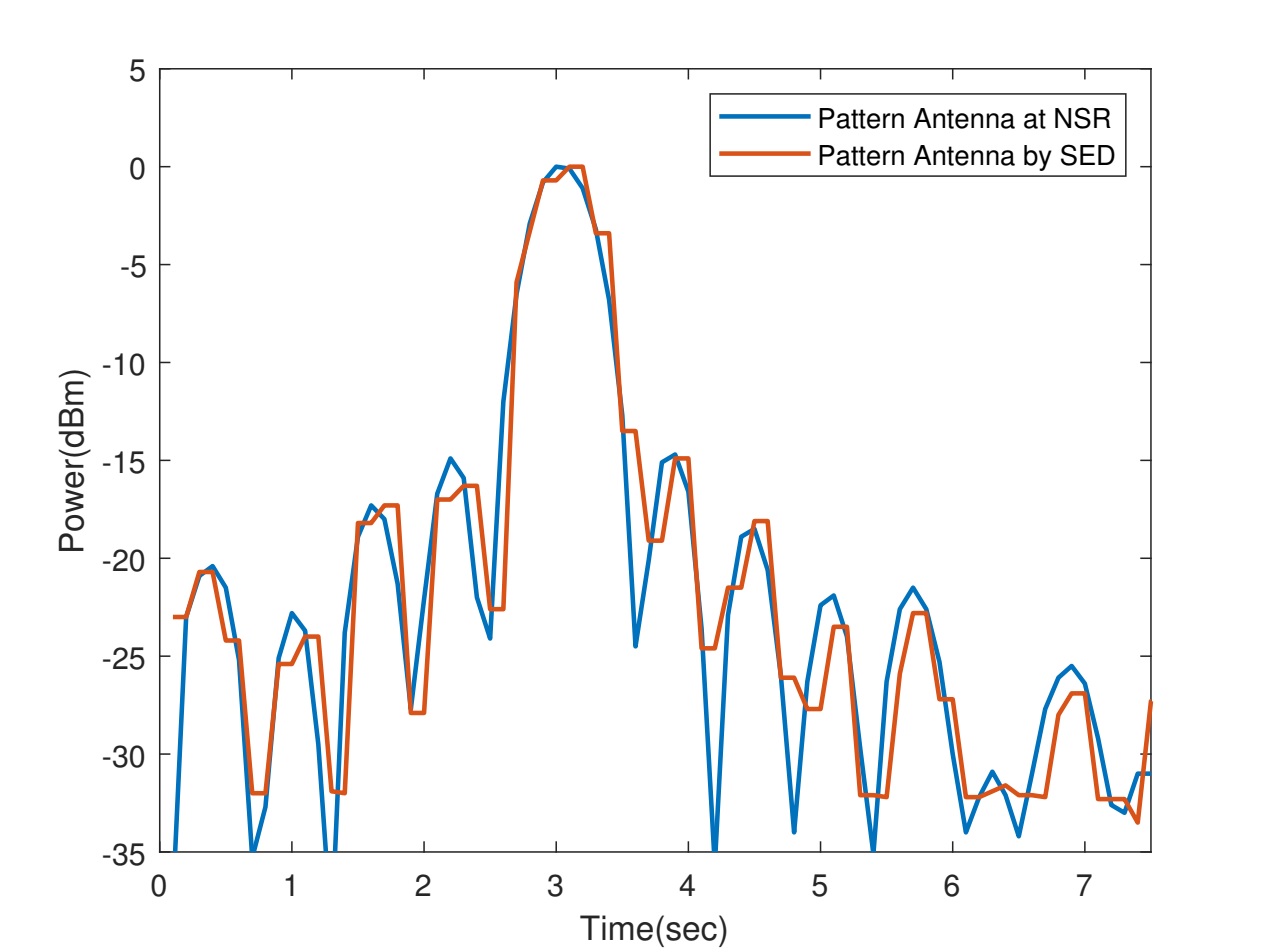}
\caption{Comparing antenna pattern between the simple envelope detector and NSR.}
\label{Fig8}
\end{figure}

At 14:29:22 UTC, a signal was received from Cosmo-SkyMed satellite for 7 seconds. Fig. \ref{Fig6} illustrates the spectrum of the monitored signal at different times. The sweep time of the receiver is set to 100 ms, so data points at each frequency are recorded at a rate of 10 points per second. By compiling the power data across frequencies, an antenna pattern can be drawn for each frequency over time. Additionally, it can be plotted with respect to the angle if the ground speed of the antenna beam and the distance between the satellite and the receiver are known.
Fig.\ref{Fig7} illustrates the antenna pattern at frequencies of 9551 MHz and 9614.5 MHz. According to the theoretical results in Section 2, the distance between the nulls increases proportionally with the wavelength.
Although the distance between the frequencies was consistent at to 63.5 MHz, the antenna patterns did not match after the fourth second. The mismatch increased as the frequency distance increased. Fig.\ref{Fig7} demonstrates the improved performance of the receiver, which can obtain the antenna pattern independently for each frequency.
In this section, the antenna patterns obtained using a simple envelope detector and NSR are compared. The NSR is better at extracting the antenna pattern than the simple envelope detector, because the bandwidth of SED was relatively wide at 400 MHz while the bandwidth of NSR was 10 MHz. Fig. \ref{Fig8} validates, the receiver's performance in terms of obtaining the antenna pattern with greater accuracy than the simple envelope detector was confirmed.
\section{Conclusion}
This paper introduced a novel Narrowband Sweeper Receiver (NSR) for precise estimation of antenna patterns in SAR-equipped satellites. The NSR effectively measured power at individual frequencies, capturing key antenna spectrum characteristics, including shape, central frequency, and bandwidth. By isolating antenna patterns for each frequency, the NSR enhanced analytical precision compared to conventional receivers. Analytical and simulation results confirmed that the distance between nulls increases with wavelength, a finding supported by practical tests. Additionally, the narrowband filtering mechanism significantly improved the signal-to-noise ratio (SNR), validated through theoretical and experimental analyses.

\end{document}